\documentclass[aps,twocolumn,showpacs,prl,superscriptaddress]{revtex4-2}
\usepackage{mathtools}
\usepackage{bm}
\usepackage{dsfont,amsthm,amsbsy}
\usepackage{verbatim}
\usepackage{amssymb}
\usepackage{amsmath}
\usepackage{bbm}
\usepackage{graphicx}
\usepackage{epstopdf}
\usepackage{subfigure}
\usepackage{epsfig}
\usepackage{amsfonts}
\usepackage{mathrsfs}
\usepackage{sidecap}
\usepackage{lipsum}
\usepackage[toc,page,title,titletoc,header]{appendix}
\usepackage[colorlinks,linkcolor=blue,citecolor=blue,anchorcolor=blue,urlcolor=blue]{hyperref}
\usepackage{hyperref}
\usepackage{resizegather}
\usepackage{tikz}
\usepackage{float}
\usepackage{mathbbol}
\usepackage[normalem]{ulem}
\usepackage{cancel}
\usepackage{upgreek}

\newcommand{\bl}{\begin{aligned}}
\newcommand{\el}{\end{aligned}}

\def\be{\begin{equation}}
\def\ee{\end{equation}}

\def\bi{\begin{itemize}}
\def\ei{\end{itemize}}
\def\bn{\begin{enumerate}}
\def\en{\end{enumerate}}
\def\bea{\begin{eqnarray}}
\def\eea{\end{eqnarray}}
\def\no{\nonumber}
\def\ba{\begin{array}}
\def\ea{\end{array}}
\def\bd{\begin{displaymath}}
\def\ed{\end{displaymath}}

\begin{document}
\title
{Dynamical quantum phase transitions following a noisy quench}


\author{R. Jafari}
\email[]{jafari@iasbs.ac.ir, raadmehr.jafari@gmail.com}
\affiliation{Department of Physics, Institute for Advanced Studies in Basic Sciences (IASBS), Zanjan 45137-66731, Iran}
\affiliation{School of Nano Science, Institute for Research in Fundamental Sciences (IPM), 19395-5531, Tehran, Iran}
\affiliation{Department of Physics, University of Gothenburg, SE 412 96 Gothenburg, Sweden}

\author{A. Langari}
\email[]{langari@sharif.edu}
\affiliation{Department of Physics, Sharif University of Technology, P.O.Box 11155-9161, Tehran, Iran}

\author{S. Eggert}
\email[]{eggert@physik.uni-kl.de}
\affiliation{Physics Department and Research Center OPTIMAS,
University of Kaiserslautern, 67663 Kaiserslautern, Germany}

\author{Henrik Johannesson}
\email[]{henrik.johannesson@physics.gu.se}
\affiliation{Department of Physics, University of Gothenburg, SE 412 96 Gothenburg, Sweden}

\date{\today}

\begin{abstract}
We study how time-dependent energy fluctuations impact the dynamical quantum phase transitions (DQPTs) following a noisy ramped quench of the transverse magnetic field in a quantum Ising chain. By numerically solving
{\color{black} the stochastic Schr\"odinger equation of the mode-decoupled fermionic Hamiltonian of the problem, we identify two generic scenarios: Depending on the amplitude of the noise and the rate of the ramp, the expected periodic sequence of noiseless DQPTs may either be uniformly shifted in time or else replaced by a disarray of closely spaced DQPTs. Guided by an exact noise master equation, we trace the phenomenon to the interplay between noise-induced excitations which accumulate during the quench and the near-adiabatic dynamics of the massive modes of the system. Our analysis generalizes to any 1D fermionic two-band model subject to a noisy quench.}

\end{abstract}

\maketitle
Dynamical quantum phase transitions (DQPTs) have become one of the focal points in the study of quantum matter out of equilibrium \cite{Heyl2018,Zvyagin2016}, spurred by the prospect of performing high-precision tests using quantum simulators \cite{Georgescu2014,Altmann2021}. DQPTs appear at critical times at which the overlaps between initial and time-evolved states vanish.
{\color{black} As a result, the rate function
which plays the role of a dynamical free energy density \cite{Heyl2013} becomes nonanalytic in the thermodynamic limit.
{\color{black} With} time replacing the usual notion of a control parameter, DQPTs are different from ordinary phase transitions, requiring new ideas and concepts for their understanding. Progress has come thick and fast, with} an expanding literature on theory \cite{Heyl2013,Karrasch2013,Kriel2014,Canovi2014,Andraschko2014,vajna2014disentangling,Heyl2014,Heyl2015,vajna2015topological,Budich2016,
Sharma2016b,Huang2016,Bhattacharya2017,Heyl2017,bhattacharya2017emergent,zhou2018dynamical,Lang2018,Zhou2019,Mendl2019,Huang2019,
Uhrich2020,Nicola2021,Bandyopadhyay2021,Jafari2021,Halimeh2021,Sacramento2021,Peotta2021,Damme2022,Hou2022}, modeling, and experimentation \cite{jurcevic2017direct,Zhang2017,Bernien2017,flaschner2018observation,guo2019observation,wang2019simulating,YangTianYang2019,Tian2019,Tian2020,Chen2020,Nie2020,Xu2020,Zhao2021,
WuNettersheim2022}.

Most research so far, theoretical as well as experimental, has considered DQPTs triggered by a quantum quench where an isolated system is forced out of equilibrium by a change of its Hamiltonian. The quench may be modeled as sudden, or more realistically, as having a finite duration with a Hamiltonian parameter being swept from an initial to a final value, also known as  a ``ramp". While the quench is usually assumed to be governed by a well-defined Hamiltonian, its realization in an experiment is always imperfect. As a result, when energy is transferred into or out of an otherwise isolated system via a quench in the laboratory, there will inevitably be time-dependent fluctuations (``noise") in this transfer. Examples include noise-induced heating caused by amplitude fluctuations of the lasers forming an optical lattice \cite{Doria2011} and fluctuations in the effective magnetic field applied to a system of trapped ions \cite{Britton2012}. This raises the important issue about the robustness of DQPTs following a {\em {\color{black} noisy}} quench. Do the DQPTs survive? If so, what is the effect from noise on the dynamical critical behavior?

We address these questions in the setting of the transverse field Ising (TFI) chain, arguably the simplest benchmark model for this purpose. {\color{black} The model has served as a paradigm for exploring quantum phase transitions in and out of equilibrium, and is also the first \cite{Heyl2013} and best studied model exhibiting DQPTs. The availability of platforms for well-controlled experimental probes of DQPTs in TFI-like chains \cite{jurcevic2017direct,Zhang2017,Nie2020,Bernien2017,Chen2020} is yet another reason why we choose it for our study. Quantitative reliable results for the simple TFI chain, amenable to experimental tests, should prepare the ground for a comprehensive theory of DQPTs following a noisy quench.}

Representing the noise by a dynamical stochastic variable added to the TFI Hamiltonian, {\color{black} we numerically study the stochastic Schr\"odinger equation of the corresponding mode-decoupled fermionic Hamiltonian that governs the dynamics of a single quench. In addition, we construct and solve an exact master equation for the quench dynamics averaged over the noise distribution.} This allows us to highlight the interplay between the near-adiabatic quench dynamics of the gapped modes of the system and the accumulation of noise-induced excitations. {\color{black} As suggested by our analysis, the competition between adiabaticity and noise-induced excitations underlies the sometimes surprising outcome of a noisy quench.} While a small ratio between noise amplitude and rate of energy transfer {\color{black} at most} results in a shift of the {\color{black} expected periodic sequence of noiseless} DQPTs, a larger ratio may have a dramatic effect: {\color{black} The periodic sequence can now get scrambled, resulting in a disarray of closely spaced DQPTs.}

To set the stage, we write down the Hamiltonian of the Ising chain with periodic boundary conditions and subject to a noiseless transverse magnetic field $h_0(t)$,
%
\begin{equation}
H_{0}(t) =- J\sum_{n=1}^{N} \sigma_{n}^{x} \sigma_{n+1}^{x}
- h_0(t) \sum_{n=1}^{N} \sigma_{n}^{z}.
\label{eq:Ising}
\end{equation}
%
When the field is time-independent, $h_0(t) = h$, and with $J$ set to unity, the ground state is ferromagnetic for $|h| < 1$, otherwise paramagnetic, with the phases separated by equilibrium quantum critical points at $h=\pm 1$ \cite{Pfeuty1970}. Here, and in what follows, {\color{black} $\hbar=1$.}

The Hamiltonian $H_0(t)$ in Eq. (\ref{eq:Ising}) can be mapped onto a model of spinless fermions with operators $c_n, c_n^\dagger$ using a Jordan-Wigner transformation \cite{LSM1961}. Performing a Fourier transformation, $c_n = (\mbox{e}^{i\pi/4}/\sqrt{N}) \sum_k \mbox{e}^{ikn}c_k$ (with the phase factor $\mbox{e}^{i\pi/4}$ added for convenience), $H_0(t)$ gets expressed as a sum over decoupled mode Hamiltonians $H_{0,k}(t)$,
%
\bea
\label{eq:Nambu}
H_0(t) = \sum_{k} C^{\dagger}_k H_{0,k}(t) C_k, \ \ k= \frac{(2m-1)\pi}{N},
\eea
%
with $m=1, 2, \dots, N/2$. Here $N$ and the fermion parity $\exp(i\pi\sum_{n=1}^N a_n^\dagger a_n)$ are taken to be even \cite{Franchini2017}.
$C_k^{\dagger} = (c_k^{\dagger} \ c_{-k})$ are Nambu spinors, and
%
\bea
\label{eq:BdG}
H_{0,k}(t) = h_{0,k}(t)\sigma^z + \Delta_k\sigma^x
\eea
%
with $h_{0,k}(t) = 2(h_0(t)-\cos(k))$ and $\Delta_k = 2\sin(k)$ when $J=1$. The instantaneous eigenstates and eigenvalues of $H_{0,k}(t)$ are given by
%
\bea
\label{eq:eigenstates}
|\chi_k^{\pm}(t)\rangle&=&\alpha_k^{\pm}(t) |\alpha\rangle+\beta_k^{\pm}(t)|\beta\rangle, \\
\label{eq:spectrum}
\varepsilon_k^{\pm}(t)&=&\pm\varepsilon_k(t)=\pm\sqrt{h^{2}_{0,k}(t)+\Delta_k^2},
\eea
%
where $|\alpha\rangle = (1\, 0)^T$, $|\beta\rangle = (0\, 1)^T$, and $\alpha_k^{\pm}(t) \!= \!(h_{0,k}(t)\!\mp\!\varepsilon_k(t))/N_k^{\pm}(t)$, \,$\beta_k^{\pm}(t)
\!=\! \Delta_k/N_k^{\pm}(t)$, with $N_k^{\pm}(t)$ normalization constants. Note that for large $N$, the gap between the two levels vanishes in the limit $k \rightarrow \pi \ (k \rightarrow 0)$ when reaching the critical points $h_0(t) \!=\! - 1 $ ($h_0(t) \!=\! + 1 $). Also note that the Pauli matrices in Eq.~(\ref{eq:BdG}) are not to be mixed up with the spin operators in Eq.~(\ref{eq:Ising}).

As a preliminary, let us briefly review DQPTs in case of a noiseless ramp {\color{black} with sweep velocity $v$}, $h_0(t) = h_f + vt$, from an initial value $h_i$ at time $t={t}_i<0$ to a final value $h_f$ at $t\!=\!t_f\!=\!0$.
The Hamiltonian in Eq.~(\ref{eq:BdG}) for each mode has the form
$H_{0,k}(t) = \frac{1}{2}v\tau_k \sigma_z + \Delta_k \sigma_x$, so transition rates can be
calculated  by the Landau-Zener formula \cite{Landau,Zener}.
Here $\tau_k = 2h_{0,k}(t)/v$ defines a mode-dependent time, which changes sign when an avoided level crossing  occurs \cite{Dziarmaga2005, Damski2005}.
As expected from the adiabatic theorem \cite{Kato1950}, a {\color{black} quasiparticle mode with wave number $k$} remains in its instantaneous eigenstate in the limit {\color{black} $v\Delta_k/2\varepsilon^3_k(t) \rightarrow 0$  \cite{Vitanov1999} (with $2\varepsilon_k(t)$ the gap of {\color{black} the mode} at time $t$; cf. Eq. (\ref{eq:spectrum}))}, hence $\{|\chi^{\pm}(\tau_k)\rangle\}$ span the adiabatic basis, with $\{|\alpha\rangle, |\beta\rangle \}$ the diabatic basis.

Starting with $h_i \ll -1$ in the
ground state of the paramagnetic phase,
all modes initially reside in the lower level $|\chi^{-}_k({t}_i)\rangle$.
After a ramp across the critical field $h=-1$ to some final value $h_f=1/2$ in the
ferromagnetic phase, {\color{black} the probability
to find mode $k$ in the upper level $|\chi^{+}_k({t}_f)\rangle$
will depend on the value of $k$, and we denote this probability by $p_k$.}
Modes close to $k=0$ show no sign change of $\tau_k$, so they
mostly remain in the lower level $p_k<1/2$, while modes close to the gap-closing limit $k=\pi$ will be excited to the upper level with probability $p_k>1/2$ \cite{Heyl2013,Sharma2016b}. Given these two cases, continuity of the spectrum as a function of $k$ in the thermodynamic limit implies that there exists a ``critical mode" $k^{\ast}$ with equal
probabilities $p_{k^{\ast}}=1/2$ for occupation of the lower and upper levels after the ramp,
corresponding to a maximally mixed state. This is the
mode that triggers the appearance of DQPTs at critical times \cite{Heyl2018,Jafari2022}
%
\begin{equation}
\label{eq:criticaltimes}
t_c^n =(2n+1)\frac{\pi}{2\varepsilon_{k^{\ast}\!,f}}, \ \ n=0,1,...
\end{equation}
with $\varepsilon_{k^{\ast}\!,f} = \varepsilon_{k^{\ast}}(t_f)$ the energy in Eq.~(\ref{eq:spectrum}).
Note that the ramp occurs at negative times, $t<t_f=0$, while the DQPTs take place at
positive times. \nocite{Kolodrubetz2012,Sancho1982,Fox1988,Billah1990,Mannella1992,Dagpunar1988,Novikov1965,Barouch1970,Chenu2017}


To approach the problem with a noisy quench we add a random variable $\eta(t)$ to the magnetic field, writing $h(t) = h_0(t)+ \eta(t)$. We shall assume the noise distribution to be Gaussian with vanishing mean, $\langle \eta({t})\rangle=0$,
and with canonical Ornstein-Uhlenbeck two-point correlations
 \cite{CoxMiller1965}
%
\begin{equation}
\label{eq:noise}
\langle \eta({t})\eta({t}')\rangle=\frac{\xi^{2}}{2\tau_n}e^{-|{t}-{t}'|/\tau_n}.
\end{equation}
%
Here $\tau_n$ is the noise correlation time and $\xi$ the noise amplitude {\color{black} for fixed $\tau_n$}. The frequently employed white-noise limit is obtained by letting $\tau_n \rightarrow 0$.

As before, the probabilities $p_k$ for nonadiabatic transitions will change continuously with $k$ in the thermodynamic limit, but it
is a priori unclear if the special value $p_k\!=\!1/2$ occurs at all, or maybe even for
several $k$-values.
The inequality $p_{k,\text{max}} > 1/2$ close to $k = \pi$ is ensured by the Kibble-Zurek mechanism (KZM), which predicts a breakdown of adiabaticity when approaching gap closing \cite{Kibble1976,Zurek1985}.
On the other hand, noise will in general facilitate additional transitions, so it is uncertain if modes with $p_{k,\text{min}} < 1/2$ remain, which is the
required condition for DQPTs \cite{Heyl2018,Jafari2022}.
While there are closed expressions for finite-time transition probabilities in the adiabatic basis with no noise \cite{Vitanov1999}, there are no known such results when noise is present. Could it be that noise may increase the probability for nonadiabatic transitions, corrupting the inequality $p_{k,\text{min}} < 1/2$? {\color{black} Or maybe instead drive oscillations of the $p_k$ function across 1/2, causing additional DQPTs?}
%
\begin{figure*}[t]
\begin{minipage}{\linewidth}
\centerline{
\includegraphics[width=0.33\linewidth]{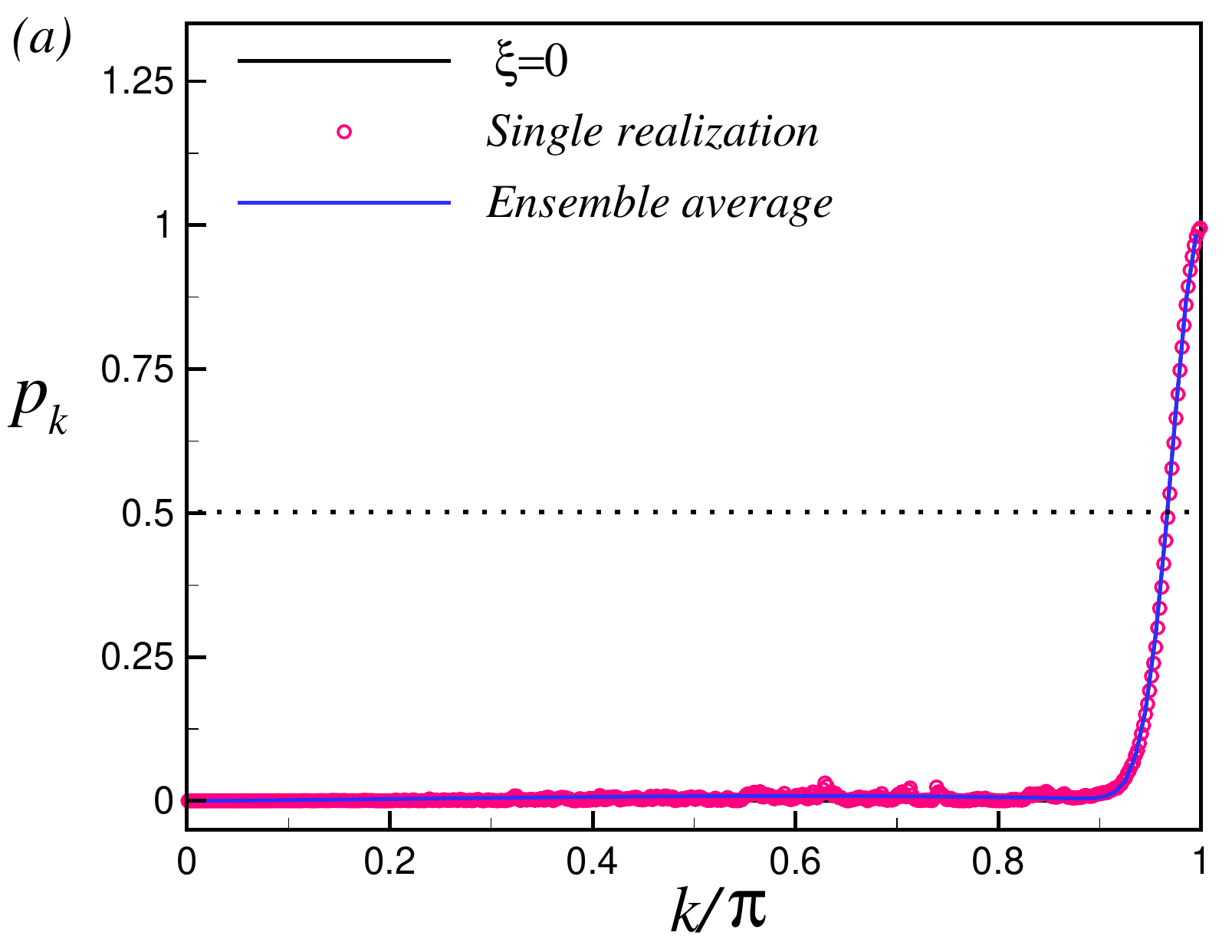}
\includegraphics[width=0.33\linewidth]{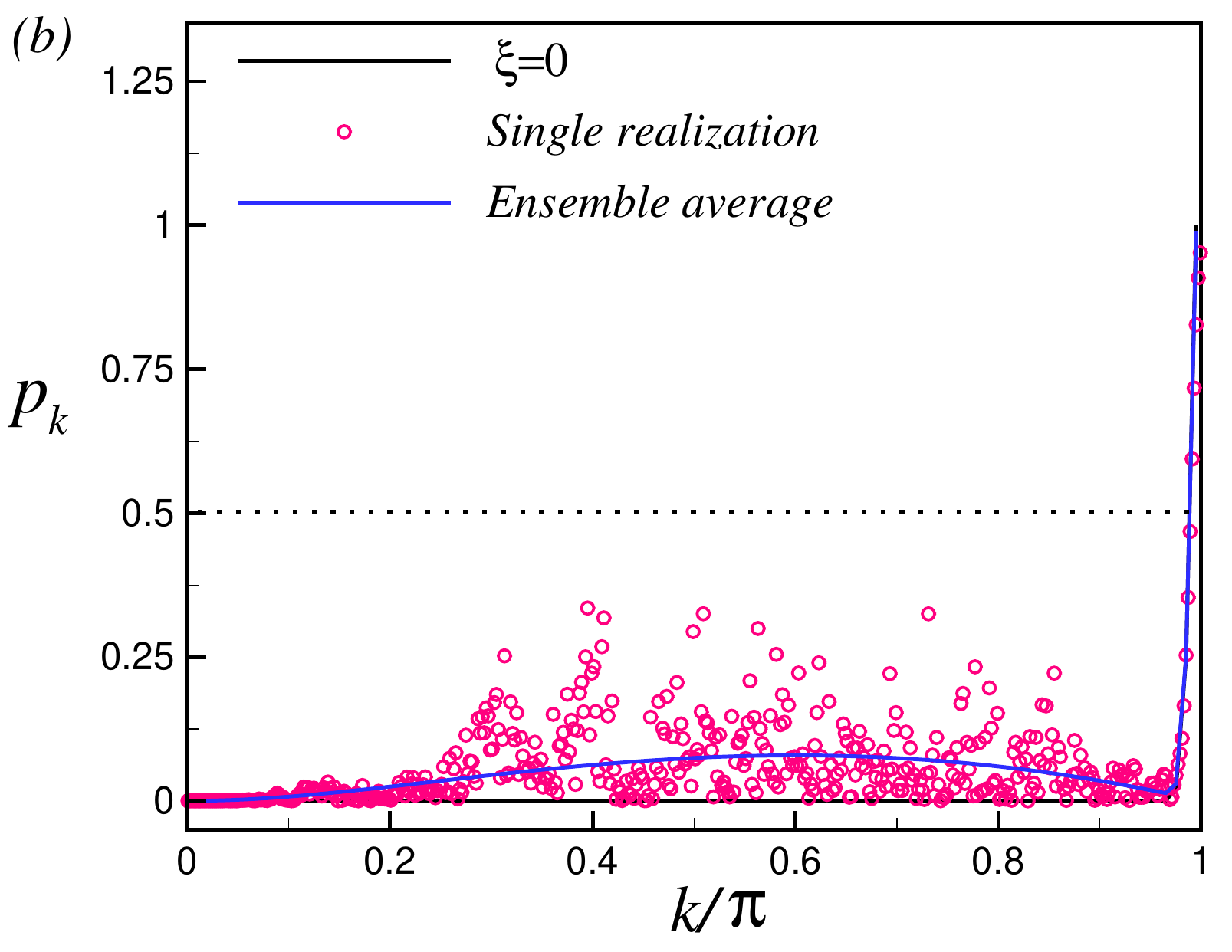}
\includegraphics[width=0.33\linewidth]{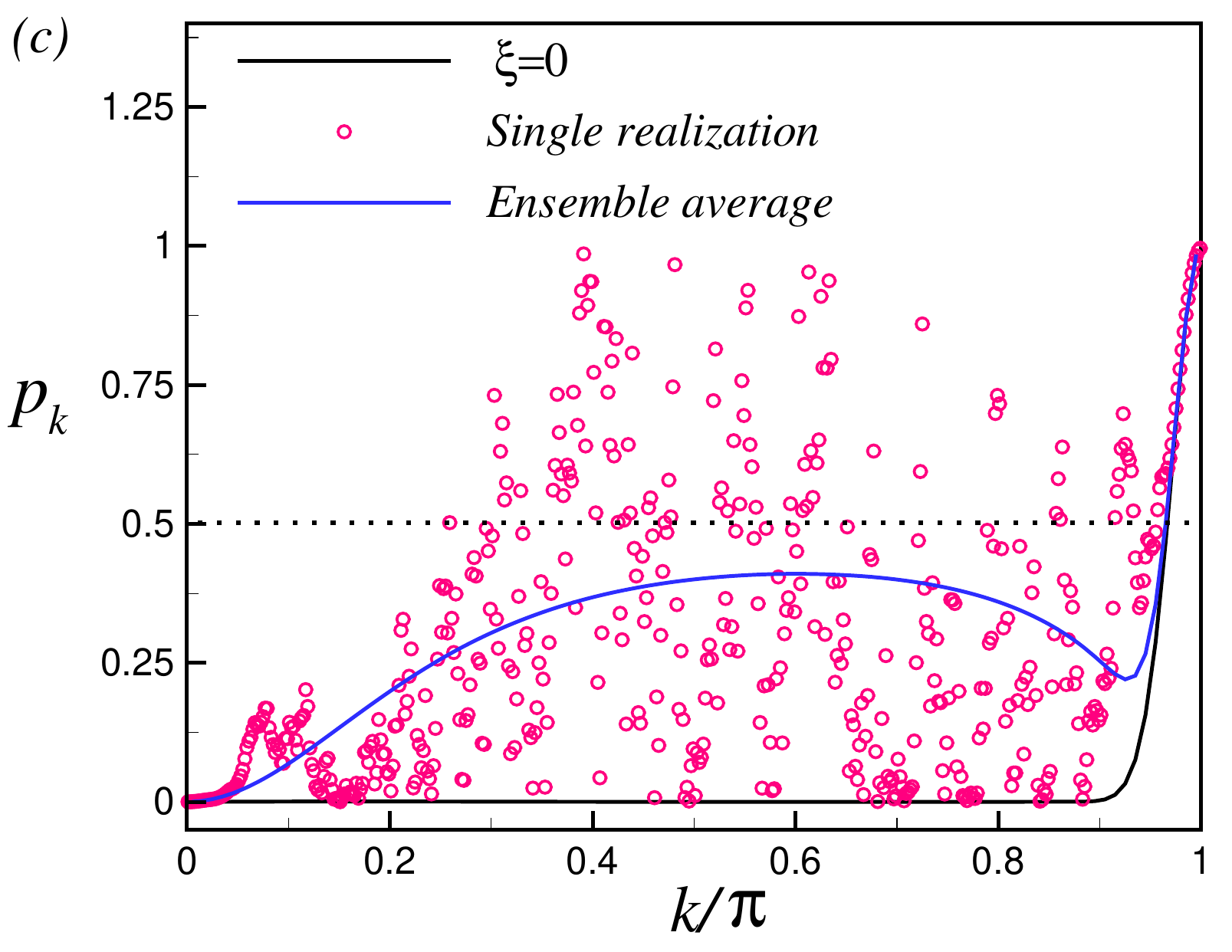}}
\centering
\end{minipage}
\begin{minipage}{\linewidth}
\centerline{
\includegraphics[width=0.33\linewidth]{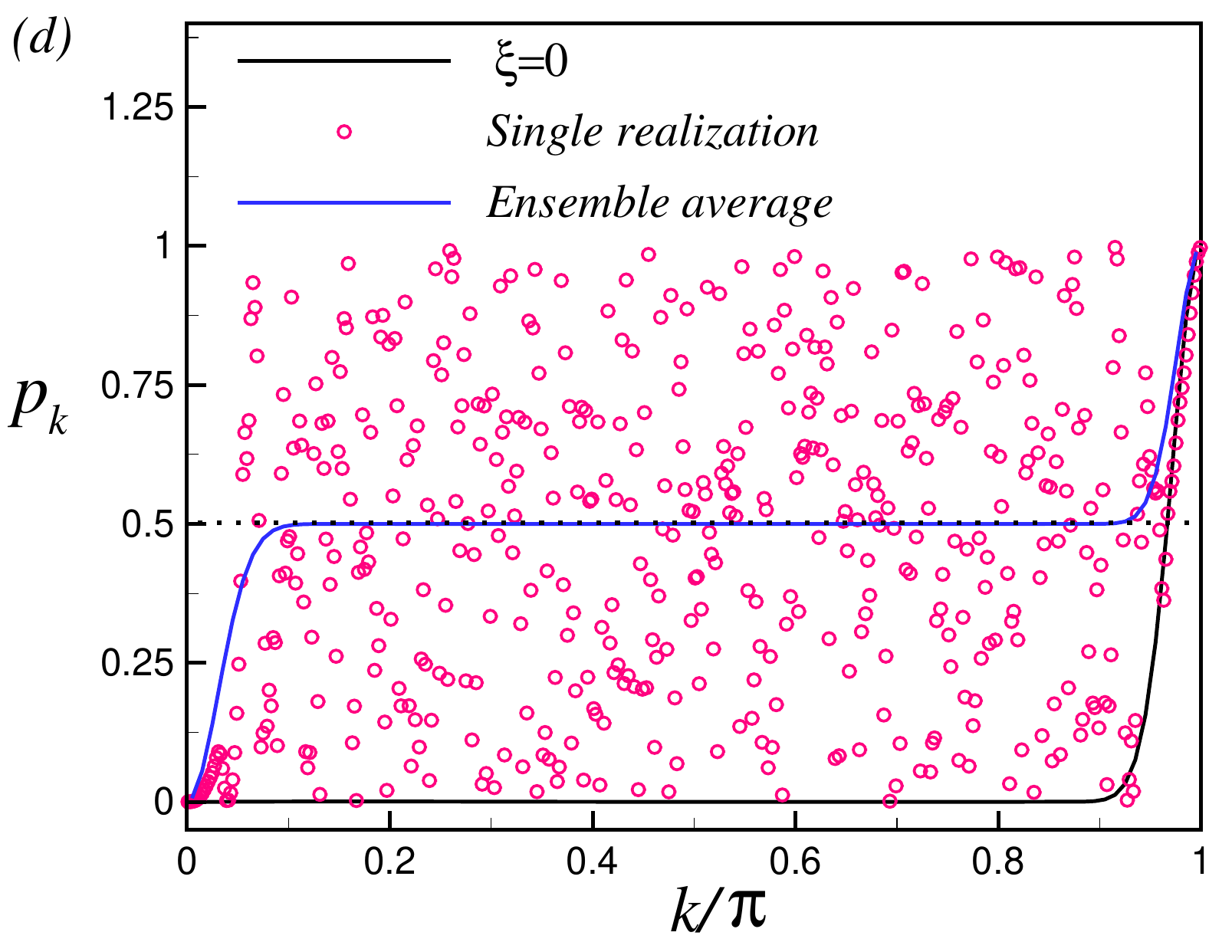}
\includegraphics[width=0.33\linewidth]{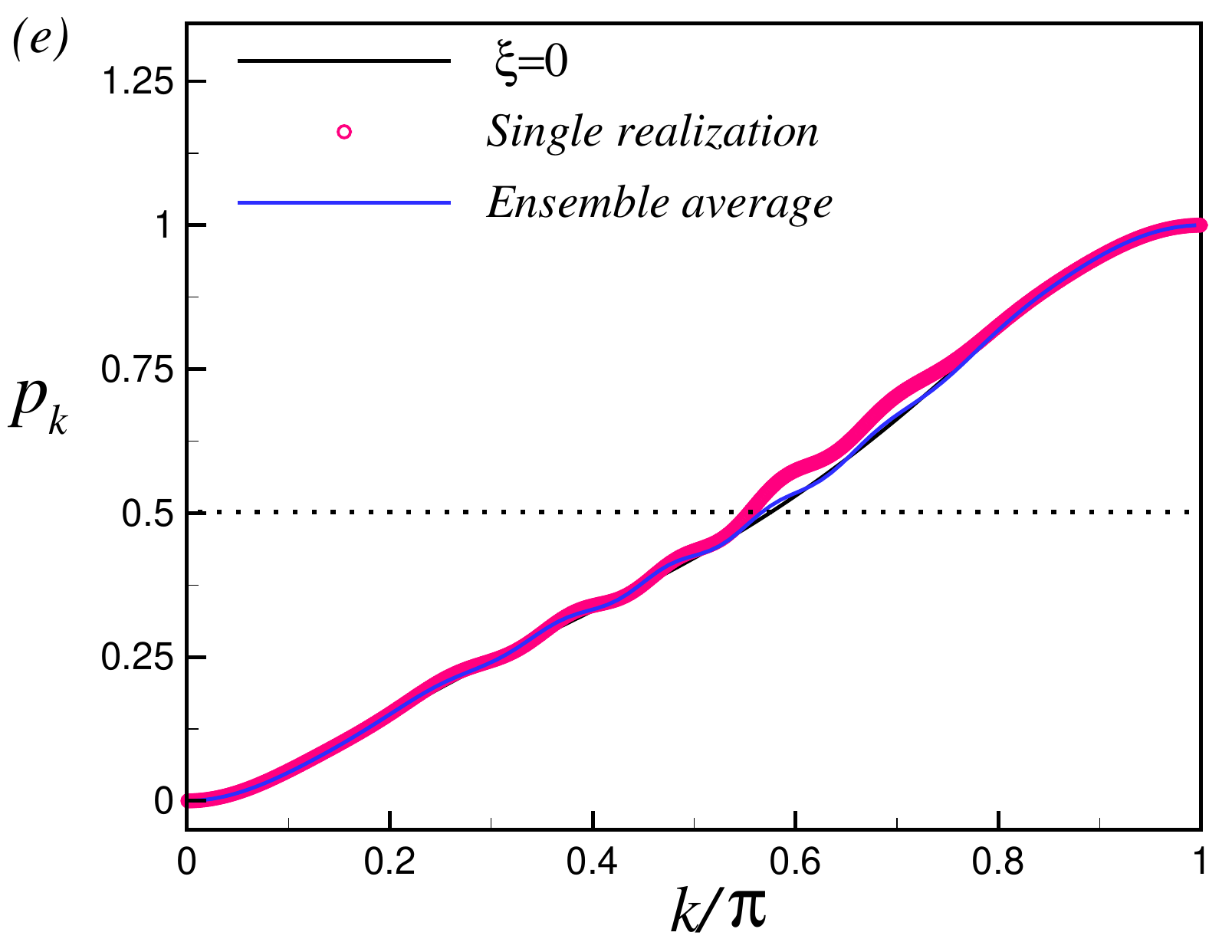}
\includegraphics[width=0.33\linewidth]{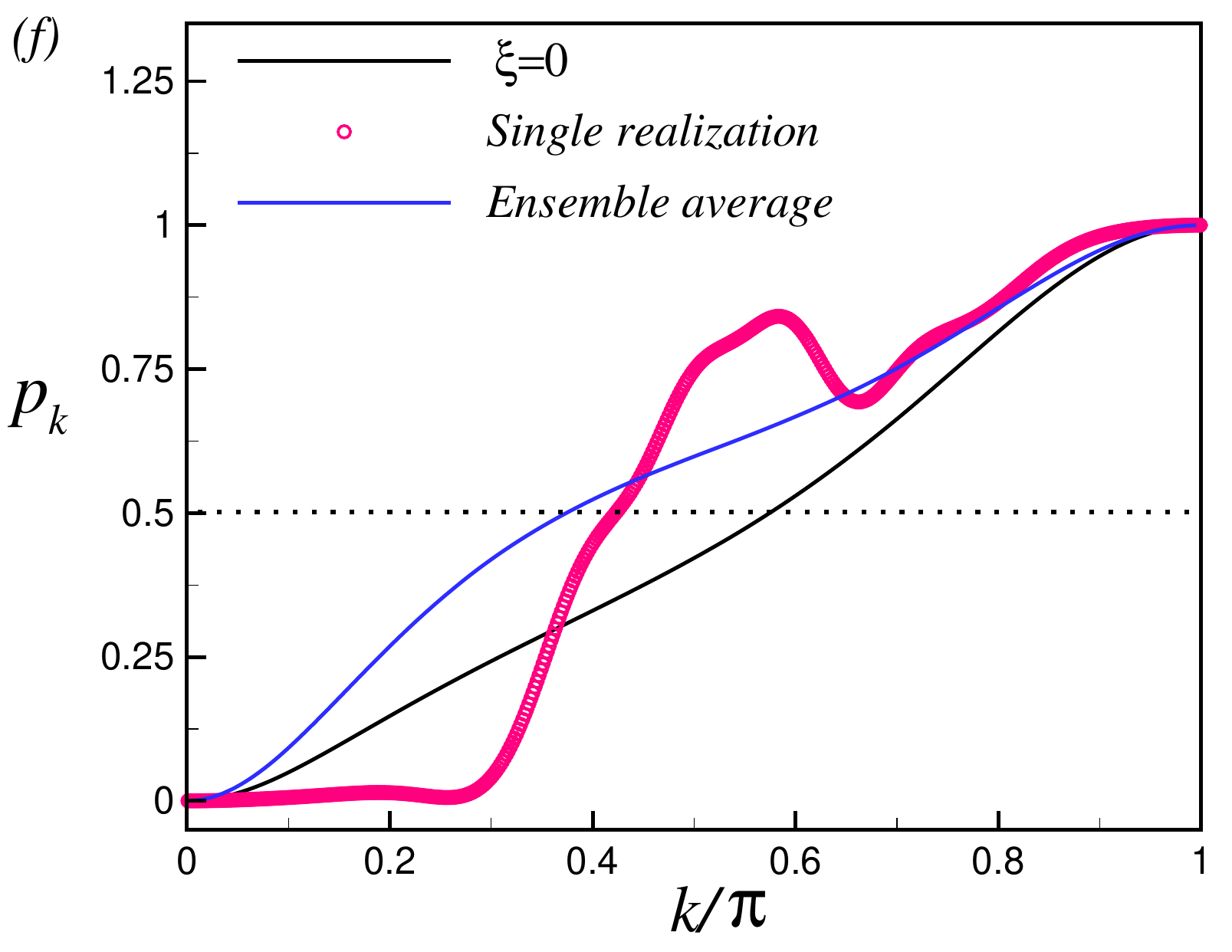}}
\centering
\end{minipage}
\caption{(Color online) {\color{black} Probabilities {\color{black} for finding a mode} with
momentum $k$ in the upper level after a ramp across the single quantum critical
point $h_c=-1$ ($h_i=-50, h_f=1/2$) for system size {\color{black} $N=1000$ and different} noise amplitudes $\xi$, sweep velocities $v$, and
noise correlation times $\tau_{n}$: (a) $\xi=0.01$, $v=0.1$, $\tau_{n}=0.01$; (b) $\xi=0.01$, $v=0.01$, $\tau_{n}=0.01$; (c) $\xi=0.1$, $v=0.1$, $\tau_{n}=0.01$;  (d) $\xi=1$, $v=0.1$, $\tau_{n}=0.01$; (e) $\xi=0.1$, $v=10$, $\tau_{n}=0.01$; (f) $\xi=1$, $v=10$, $\tau_{n}=0.01$.} {\color{black} The probabilities $p_k$ for single noise realizations are displayed in red, with the ensemble averages $\langle p_k \rangle$ in blue. For comparison, the probabilities $p_k$ for noiseless cases ($\xi=0$) are shown in black.}}
\label{fig1}
\end{figure*}
%

{\color{black} To find out, we numerically solve the stochastic Schr\"odinger equations (SSEs) \cite{Gisin1992,Gaspard1999,Bouten2004}
\begin{equation} \label{eq:SSE}
\left(H_{0,k}(t) + \eta(t)H_1\right) |\psi_k(t)\rangle = i\frac{\partial}{\partial t} |\psi_k(t)\rangle
\end{equation}
{\color{black}for the allowed values of $k$ (cf. Eq. (\ref{eq:Nambu})) and for single realizations} of the noise function $\eta(t)$ in the quench interval $t \!\in \! [t_i,0]$, with $H_1\!= \!2\sigma^{z}$ (cf. Eq. (\ref{eq:BdG}) with $h_{0,k}(t) \rightarrow h_{0,k}(t) + \eta(t)$). Having obtained the solution
$|\psi_k(t)\rangle = u_k(t)|\chi_k^+(t)\rangle +v_k(t)|\chi_k^-(t)\rangle$ to Eq. (\ref{eq:SSE}) at $t\!=\!t_f\!=\!0$, one reads off $p_k = |u_k(0)|^2$ \cite{Jafari2022}.

In addition we construct an exact noise master equation (ME) \cite{luczka1991quantum,Budini2000,Filho2017,Kiely2021} for the averaged density matrix $\rho_k(t) = \langle \rho_{\eta,k}(t) \rangle$,
with $\rho_{\eta,k}(t)$ the density matrix of the Hamiltonian in Eq. (\ref{eq:SSE}).} Explicitly,
%
\begin{eqnarray}
\label{eq:master}
\dot{\rho}_k(t)=&&-i[H_{0,k}(t),\rho_k(t)]\\
\no
&&-\frac{\xi^{2}}{2\tau_n}\Big[H_{1},\int_{t_i}^{t}e^{-|t-s|/\tau_{n}}[H_{1},\rho_k(s)]ds\Big].
\end{eqnarray}
%
By translating Eq. (\ref{eq:master}) into two coupled differential equations, {\color{black} the mean transition probabilities are obtained numerically as ensemble averages $\langle p_k\rangle$ over the noise distribution $\{\eta\}$. The averaged probabilities reveal features not easily seen from a single quench, and, moreover, {\color{black} allows us to} validate the soundness of the SSE numerics.} For details, see \cite{Jafari2022}.

{\color{black} Let us analyze the results predicted by Eqs. (\ref{eq:SSE}) and (\ref{eq:master}) for a quench across the equilibrium critical point $h=-1$, from $h_i =-50$ to $h_f=1/2$.
The effect of noise is bound to increase with the amplitude $\xi$, but will also depend on the correlation time $\tau_n$ as well as on the sweep velocity $v$. For transparency we focus on
a few representative cases, {\color{black} displayed in panels (a)-(f) of Fig. 1.}

{\color{black} (a) $-$  We take off from} a noiseless quench {\color{black} that supports an extended} adiabatic regime, i.e., with modes satisfying $p_k \!\approx \!0$. As discussed above, when a quench is noiseless there appears only a single critical momentum $k^{\ast}$ (satisfying $p_{k^{\ast}} \!= \!1/2$). {\color{black} Panel (a) shows that adding noise} in the velocity-weighted low-amplitude limit $\xi/v \ll 1$ does not perturb $k^{\ast}$. Hence, in this limit the {\color{black} corresponding} DQPTs are robust against noise.

{\color{black} (b) $-$ Increasing $\xi/v$ by lowering the sweep velocity $v$ as compared to (a)}, one enters a crossover region with $\xi/v \sim \mathcal{O}(1)$. In this region the impact of noise depends on its non-weighted amplitude $\xi$. {\color{black} Panel (b) shows that the noiseless critical momentum remains unperturbed for a sufficiently small $\xi$ (here with the same value as in (a)). Thus, the corresponding DQPTs stay robust against noise.}

{\color{black} (c) $-$ Boosting the amplitude $\xi$ in the crossover region $\xi/v \sim \mathcal{O}(1)$ (here by a factor of 10 compared to (b)) causes the $p_k$-function for a single noise realization to cross the value 1/2 for several $k$-values. The convergence of $p_k$ to a continuous function of $k$ in the thermodynamic limit $N \rightarrow \infty$ is now extremely slow, reflecting that the large-amplitude noise variability morphs into a finite-$N$ $p_k$-function with occasional large jumps between neighboring modes.
Going to larger values of $N$ will eventually smoothen the graph, implying a set of randomly distributed critical momenta $\{k^{\ast}_i\}$ in the thermodynamic limit where $p_k$ becomes continuous. By inserting $\{k^{\ast}_i\}$ into Eq. (\ref{eq:criticaltimes}) one obtains an {\em aperiodic} sequence of densely spaced DQPTs. Fig. 2 shows how such DQPTs are signaled by cusps in the dynamical free energy $g(t) = (1/N)\ln |{\cal G}(t)|$}, being finite-size precursors of the nonanalyticities in the thermodynamic limit. {\color{black} Here 
\begin{equation} \label{Loschmidt}
{\color{black} {\cal G}(t) = \prod_k \langle \psi_k(0) | \exp(-iH_{0,k}(0) t) | \psi_k(0) \rangle}
\end{equation}
is the Loschmidt amplitude for the time-evolved postquench state \cite{Jafari2022}.}

%
\begin{figure}[t]
\begin{minipage}{\columnwidth}
\centerline{
\includegraphics[width=\columnwidth]{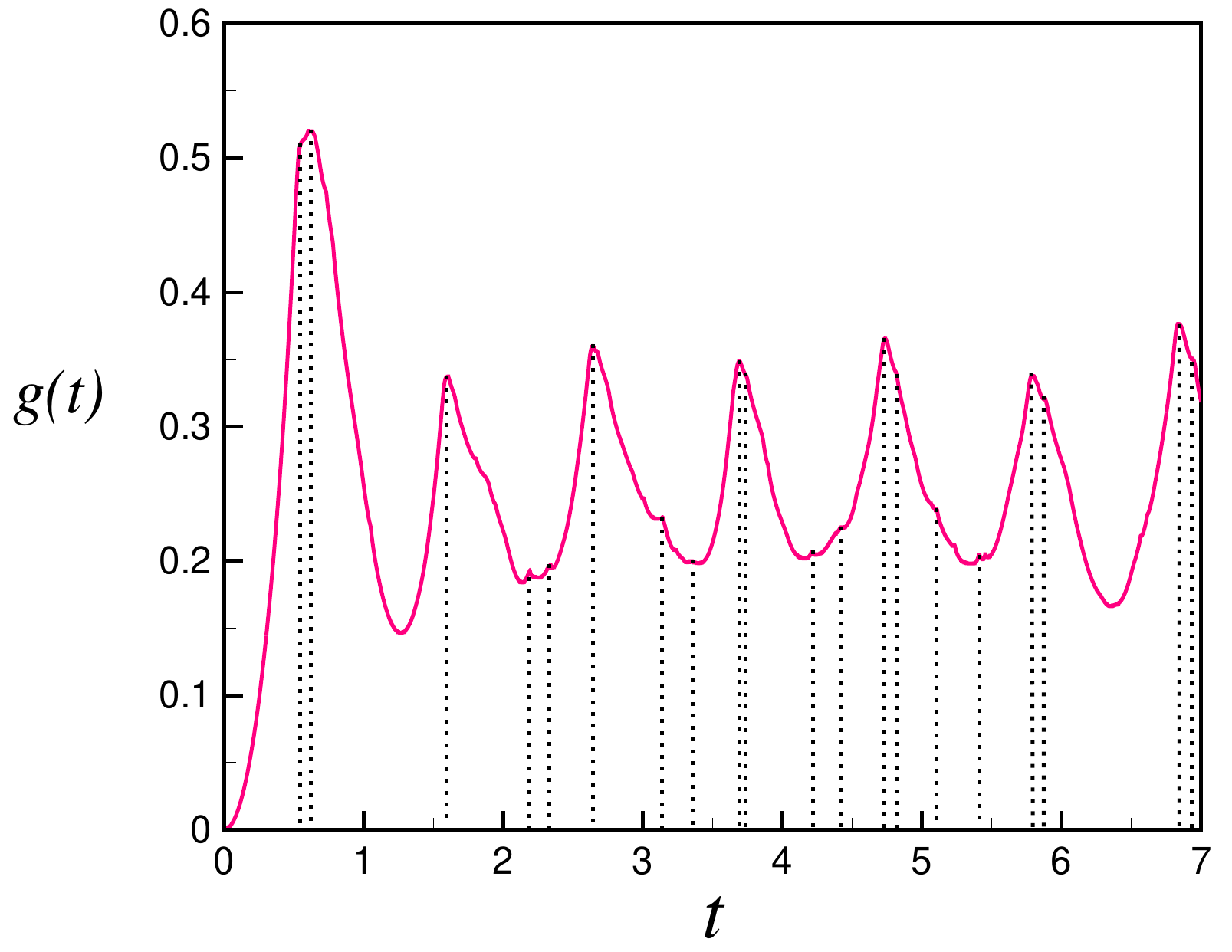}}
\caption{(Color online) The dynamical free energy $g(t)$ of the model for the noisy quench corresponding to Fig. \ref{fig1}(c). {\color{black} The vertical dotted lines mark the times for the finite-size ($N=1000$) precursors of DQPTs.}}
\label{fig2}
\centering
\end{minipage}
\end{figure}
%

{\color{black} As seen in {\color{black} both panel (b) and (c)}, the blue graphs for the ensemble averaged transition probabilities $\langle p_k\rangle$ are concave away from the gap closing region at $k\!=\!\pi$. This suggests an intriguing interplay between noise-induced excitations and the dynamics of the gapped modes driven by the slow noiseless ramp: Deep in the adiabatic regime where the {\color{black} averaged instantaneous} gap is large {\color{black} (with the average taken over the duration of the quench)}, noise has a negligible effect. For intermediate-sized gaps, noise excitations become effective but then level off as one approaches the neighborhood of $k\!=\!\pi$. Here the KZM takes over, dominating the non-adiabatic dynamics and making the presence of noise largely irrelevant.}

{\color{black} (d) $-$} Increasing $\xi/v$ further, entering the velocity-weighted large-amplitude regime $\xi/v\! \gg \!1$ (still with the noiseless quench supporting an adiabatic regime), the number of critical momenta {\color{black} in the thermodynamic limit proliferate. Similar to the case displayed in panel (c), this is spelled out by the finite-size plot of $p_k$ in panel (d) which exhibits repeated jumps of $p_k$ across the value 1/2.} As an aside, let us remark that the number of critical momenta increase also when the correlation time $\tau_n$ decreases: A smaller $\tau_n$ implies a larger noise variability $\xi/\tau_n$ which gets inherited by the $p_k$-function in the guise of a larger transition variability. {\color{black} Referring to the correlation time $\tau_n$, we also note that noise effects are conditioned by the inequality $\tau_n < 1/v$, with $1/v$ the ramp time.}

The most striking feature in {\color{black}panel (d)} is the plateau formation of the blue curve. Here the average transition probabilities $\langle p_k \rangle$ are numerically found to be locked to the value $0.5000 \pm 0.00001$, signaling the emergence of a maximally mixed state for the corresponding modes. One may understand this by noting that an Ornstein-Uhlenbeck process is stationary and therefore ergodic in the mean \cite{CoxMiller1965}. It follows that the long-time average of the noisy density matrix converges to that of its ensemble average. Given this, the formation of a plateau suggests that an asymptotically slow noisy quench will {\color{black} effectively} heat the system to infinite} {\color{black} temperature. This is supported by earlier results showing that a slow quench subject to large-amplitude white noise may lead to a maximally mixed state \cite{Pokrovsky2003,Fubini2007}. We should} add that the width of a plateau increases with decreasing $\tau_n$ as well as with decreasing $v$.

{\color{black} (e) $-$ Let us finally consider a noiseless quench where, differently from the cases (a)-(d), the assumption $v\Delta_k/2\varepsilon^3_k(t) \ll 1$  is violated for most of the modes, implying that their dynamics is nonadiabatic.
The nonadiabaticity is here driven by a larger value of the sweep velocity $v$, also giving less time for noise to become effective. As expected, and similar to the case in (a) where $\xi/v \ll 1$, panel (e) confirms that the presence of noise also now has a negligible effect when $\xi/v$ is sufficiently small.

(f)  $-$ In contrast, when $\xi/v$ is above some threshold value, however still with $\xi/v \ll 1$, the noise may {\color{black} cause a noticeable shift of} the single noiseless critical momentum, as displayed in panel (f). {\color{black} This results in a uniform} shift of the sequence of noiseless periodic DQPTs; cf. Eq. (\ref{eq:criticaltimes}).}


{\color{black} It is important to note that all DQPT scenarios in {\color{black} panels (a)-(f) of Fig. 1} are fully determined by the $p_k$ function. It follows that any 1D fermionic two-band model subject to a noisy ramp with a behavior of the $p_k$ function analogous to that of the TFI chain will show similar postquench dynamics. Let us also mention that the averaged $p_k$ curves in {\color{black}Figs. 1(a)-(f)} obtained from the ME, Eq. (\ref{eq:master}), are well reproduced by averaging over a {\color{black} finite} sample of solutions to the SSEs in Eq. (\ref{eq:SSE}), {\color{black} each SSE with a distinct noise realization $\eta(t)$;} see \cite{Jafari2022}. This serves as a stringent check on our numerical approach.

Summing up, we have shown how the patterns of DQPTs following a noisy ramped quench of the magnetic field in the TFI chain depend on the rate of the ramp (``sweep velocity" $v$) and amplitude $\xi$ of noise fluctuations. Two distinct {\color{black} classes of} scenarios can be identified: (i) noise having a negligible or weak effect, at most shifting the expected sequence of noiseless DQPTs; and (ii) noise causing an aperiodic, closely spaced, set of DQPTs. {\color{black} Note that the stochastic nature of noise does not allow us to delineate (i) and (ii) by a sharp phase boundary; only for a very small [large] ratio $\xi/v$ can we predict with certainty that (i) [(ii)] materializes after a single quench.} 

While we have here exhibited (i) and (ii) with quench protocols where one of the TFI equilibrium quantum critical points is crossed, we expect the two scenarios to be generic. Specifically, we have checked this for a ramped quench across both TFI equilibrium quantum critical points \cite{Jafari2023}.}{\color{black} The competition between adiabaticity and noise-induced excitations that brings about the two scenarios are known to be at play also in impacting the Kibble-Zurek scaling of defect formation when quenching across a critical point \cite{Fubini2007,Dutta2016,Singh2021}. It would be interesting to pinpoint related phenomena driven by this same competition.}

With the rapid advances in realizing analog quantum simulators, experimental tests of our predictions may soon be within reach. {\color{black} While we have focused our theoretical analysis on the underlying dynamics after a single noisy quench, an experimental follow-up must probably settle for ensemble averages: {\color{black} Real-time} tracking of a single-shot outcome {\color{black} will most likely have to await} further advances in weak measurement {\color{black} techniques} \cite{Hatridge2013,Jordan2024}.  On the other hand, noise-averaged (strong) measurements are expected to be fully within the realm of current {\color{black} experimental methods} and will be highly informative (as suggested by the blue-colored {\color{black} graphs} in Fig. 1).} {\color{black} Ramped} magnetic quenches in the presence of amplitude-controlled noise have already been achieved with trapped ions simulating the transverse-field XY chain \cite{Ai2021}. The other backbone for an experimental exploration $-$ detection and characterization of DQPTs $-$ is also in place, as demonstrated on a variety of platforms for {\color{black} TFI-type} chains {\color{black} with finite-range interactions}: trapped ions \cite{jurcevic2017direct,Zhang2017,Nie2020}, Rydberg atoms \cite{Bernien2017}, and NV centers \cite{Chen2020}. These breakthroughs, together with recent advances in quantum-circuit computations on NISQ devices \cite{Dborin2022,Javanmard2022}, hold promise for {\color{black} exploring DQPTs following noisy quenches also in the nearest-neighbor interacting TFI chain} studied in this Letter.

\section*{Acknowledgements}

{\color{black} We are grateful to an anonymous referee for constructive critique of an early version of this work.} SE is thankful for support from the Deutsche Forschungsgemeinschaft via Project A5 in the SFB/TR185 OSCAR. HJ acknowledges support from the Swedish Research Council via grant 621-2014-5972.

\begin{widetext}

\setcounter{figure}{0}
\setcounter{equation}{0}
\setcounter{section}{0}

\newpage

\section{Supplementary material}
\renewcommand\thefigure{S\arabic{figure}}
\renewcommand\theequation{S\arabic{equation}}
\setcounter{page}{1}

\vskip 0.5 cm

In this Supplemental material we elaborate on some technical aspects of the analysis presented in the accompanying Letter \cite{Jafari2022bsm}, and also provide some background material.

\maketitle

\setcounter{figure}{0}
\setcounter{equation}{0}
\setcounter{section}{0}

\renewcommand\thefigure{S\arabic{figure}}
\renewcommand\theequation{S\arabic{equation}}

\subsection{A. Noiseless ramp across a single equilibrium quantum critical point of the quantum Ising chain}
As a backdrop to our study of dynamical quantum phase transitions (DQPTs) following a noisy ramped quench in the quantum Ising chain, we here review the basics of the simpler case when the ramp is noiseless.

Consider the single-particle fermionic Hamiltonian $H_{0,k}({t})$ which governs the modes with wave number $k$ of the Jordan-Wigner-transformed quantum Ising chain, Eq. (3) of the main text \cite{Jafari2022bsm}. During a ramped quench in the time interval $[t_i,t_{f}]$, the transverse magnetic field $h_{0}(t)$ is swept from an initial value $h_0(t_i)=h_i$ to a final value $h_0(t_{f})=h_{f}$ such that $h_0({t}) = h_f + v{t}$, with $v>0$ the linear rate of energy transfer (``sweep velocity") during the quench. Here $t_i<0$ and $t_{f}=0$, the latter time serving as reference time for the postquench dynamics; cf. Fig. \ref{fig:LZ} (a). (For future reference we have added noise fluctuations in the figure, depicted in grey color superposed on the red-colored linear ramp.) Rewriting $H_{0,k}(t)$ on the form of a Landau-Zener model \cite{Landausm,Zenersm}, one obtains $H_{0,k}({t}) = \frac{1}{2}v\tau_k \sigma_z + \Delta_k \sigma_x$, with $\tau_k = 4(h_0(t) -\cos(k))/v$ a mode-dependent time variable, and with $\Delta_k = 2\sin(k)$.

As a case study let us look at a ramp that crosses the equilibrium quantum critical point $h_c=-1$, from the para- to the ferromagnetic phase of the model, choosing $h_0({t}_i)\!=\!h_i \!\ll \!-1$ and $h_0(t_{f})\!=\!h_{f} \!=\!1/2$. To conform to the standard Landau-Zener formalism \cite{Landausm,Zenersm}, we imagine that the ramp starts in the infinite past, $\tau_{k,i} \!=\! -\infty$. For all practical purposes, this is a viable approximation when $h_0({t}_i)\! \ll\! -1$. In this limit $H_{0,k}({t}_{i})$ effectively becomes diagonal and hence all modes initially reside in their lower level states $|\chi_k^{-}({\tau}_{k,i})\rangle \!\approx \! |\alpha\rangle$; cf. Fig. \ref{fig:LZ} (b) and Eq. (4) in the main text \cite{Jafari2022bsm}. Differently, at the end of the ramp, ${t} = {t}_f$, the $k$:th mode $|\psi_k({\tau}_{k,f})\rangle$ is in a superposition  of the upper and lower level states, $|\psi_k({\tau}_{k,f})\rangle = u_k({\tau}_{k,f}) |\chi^+_k({\tau}_{k,f})\rangle + v_k({\tau}_{k,f}) |\chi^-_k({\tau}_{k,f})\rangle$, with $|u_k|^2 + |v_k|^2 =1$. Here $|u_k({\tau}_{k,f})|^2$ is the nonadiabatic transition rate, i.e., the probability that the $k$:th mode is found in the upper level at the end of the ramp. A comment on notation: We write $p_k=|u_k({\tau}_{k,f})|^2$ and $q_k=|v_k({\tau}_{k,f})|^2$ for the transition probabilities in the adiabatic basis, to be contrasted to $P_k$ and $Q_k$ for the transition probabilities in the more frequently used diabatic basis \cite{Zenersm}. In the asymptotic limit $\tau_{k,f} \rightarrow \infty$, one has that $p_k \rightarrow Q_k$, where $Q_k$ denotes the probability for the $k$:th mode to be found in the same diabatic state in which it was initialized; cf. Fig. \ref{fig:LZ} (b) and the paragraph after Eq. (5) in the main text \cite{Jafari2022bsm}.

Zooming in on the mode $k=\pi$, we note that the gap between lower and upper levels closes when the ramp crosses $h_c=-1$ at $\tau_{k=\pi} = 0$. The generic breakdown of adiabaticity at criticality \cite{Kibble1976sm,Zurek1985sm} makes us expect that $p_k \approx 1$ for modes in the neighborhood of $k=\pi$.  In contrast, modes close to $k=0$ retain a finite gap throughout the ramp and the adiabatic theorem \cite{Kato1950} then predicts that $p_k \approx 0$ provided that the sweep velocity $v$ is sufficiently small. One expects that the $k \approx 0$ dynamics remains predominantly adiabatic also for larger values of $v$, i.e., $p_k<1/2$, provided that the condition $h_0(t_f)=1/2$ stays put (since otherwise the $k\approx 0$ modes may experience a breakdown of adiabaticity if getting too close to the other equilibrium quantum critical point at $h\!=\!1$ where their gaps approach zero). As seen from the black curves in Fig. 1 of \cite{Jafari2022bsm}, the above-mentioned expectations are corroborated by a numerical solution of the noise master equation for the averaged density matrix of the Hamiltonian $H_{0,k}(t)$, cf. Eq. (8) in \cite{Jafari2022bsm} and subsection B of this Supplemental Material. It follows that the upper and lower bounds of $p_k$ bracket $1/2$, implying that by continuity in the thermodynamic limit there is certain to be a mode, call it $k^{\ast}$, such that $p_{k^{\ast}} = 1/2$. Such a mode will trigger a DQPT.

{\color{black} The key importance of a critical mode $k^{\ast}$ was noted already in the seminal work by Heyl {\em et al.} \cite{Heyl2013sm}. These authors realized that its existence is generically a sufficient condition for the appearance of DQPTs in systems with effective descriptions in terms of a two-band fermionic model. In short, they showed that in the thermodynamic limit the maximally mixed state coded by $p_{k}\!=\!1/2$ implies that the Fisher zeros (i.e., the zeros of the partition function analytically continued to the complex plane) coalesce to a family of lines $n=0,1,2,...$ which $-$ for a quench across an equilibrium critical point $-$ are guaranteed to cut the time axis at $t^{\ast}_n, n=0,1,2,...$. This produces the nonanalyticities that define the DQPTs at times $t^{\ast}_n$.} {\color{black} The special role of a mode with $p_k=1/2$} had also been discussed earlier by Kolodrubetz {\em et al.} \cite{Kolodrubetz2012sm}, also for the quantum Ising chain, but in a slightly different context.

%
\begin{figure*}[h]
\begin{minipage}{\linewidth}
\centerline{\includegraphics[width=0.36\linewidth,height=0.28\linewidth]{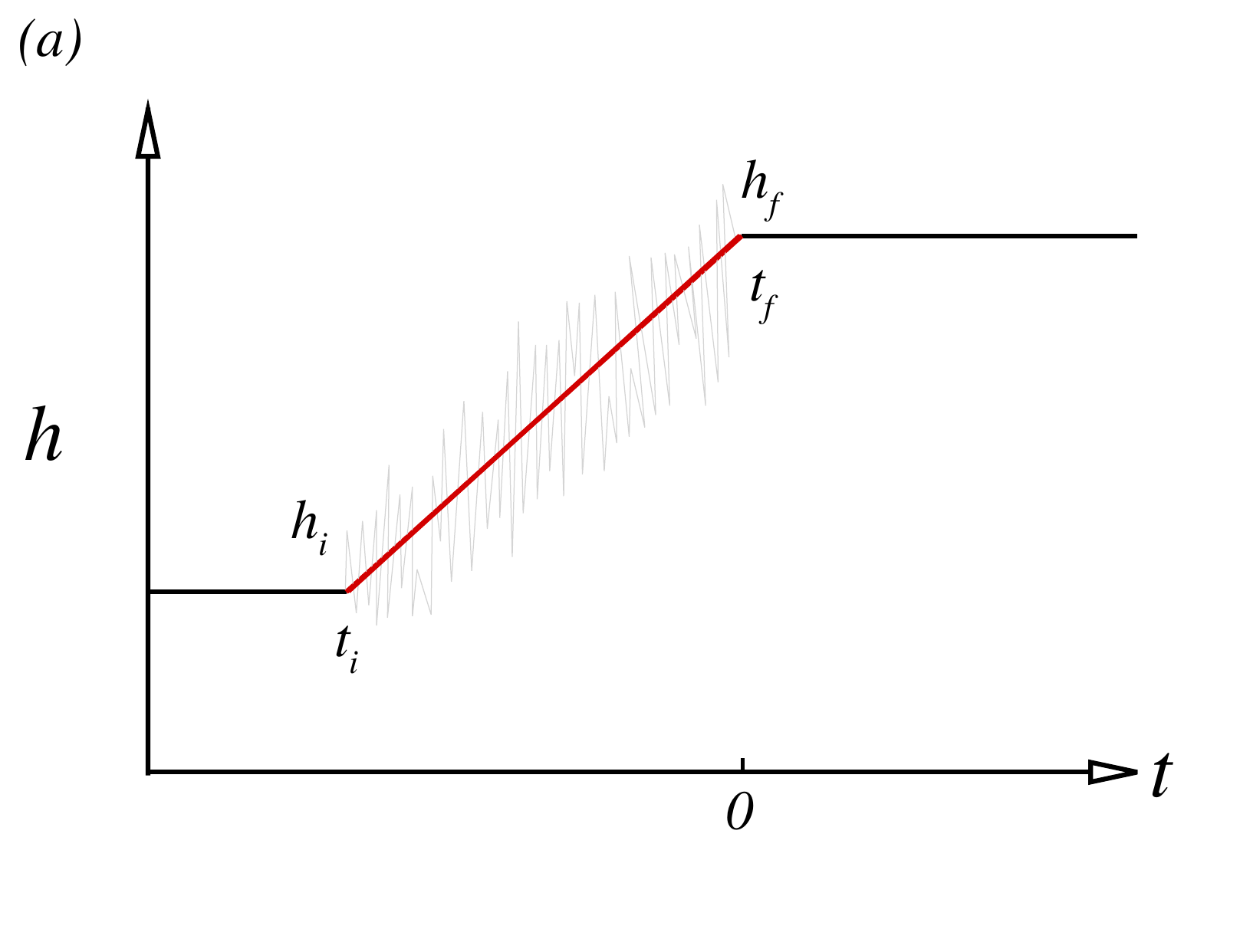}
\includegraphics[width=0.36\linewidth,height=0.28\linewidth]{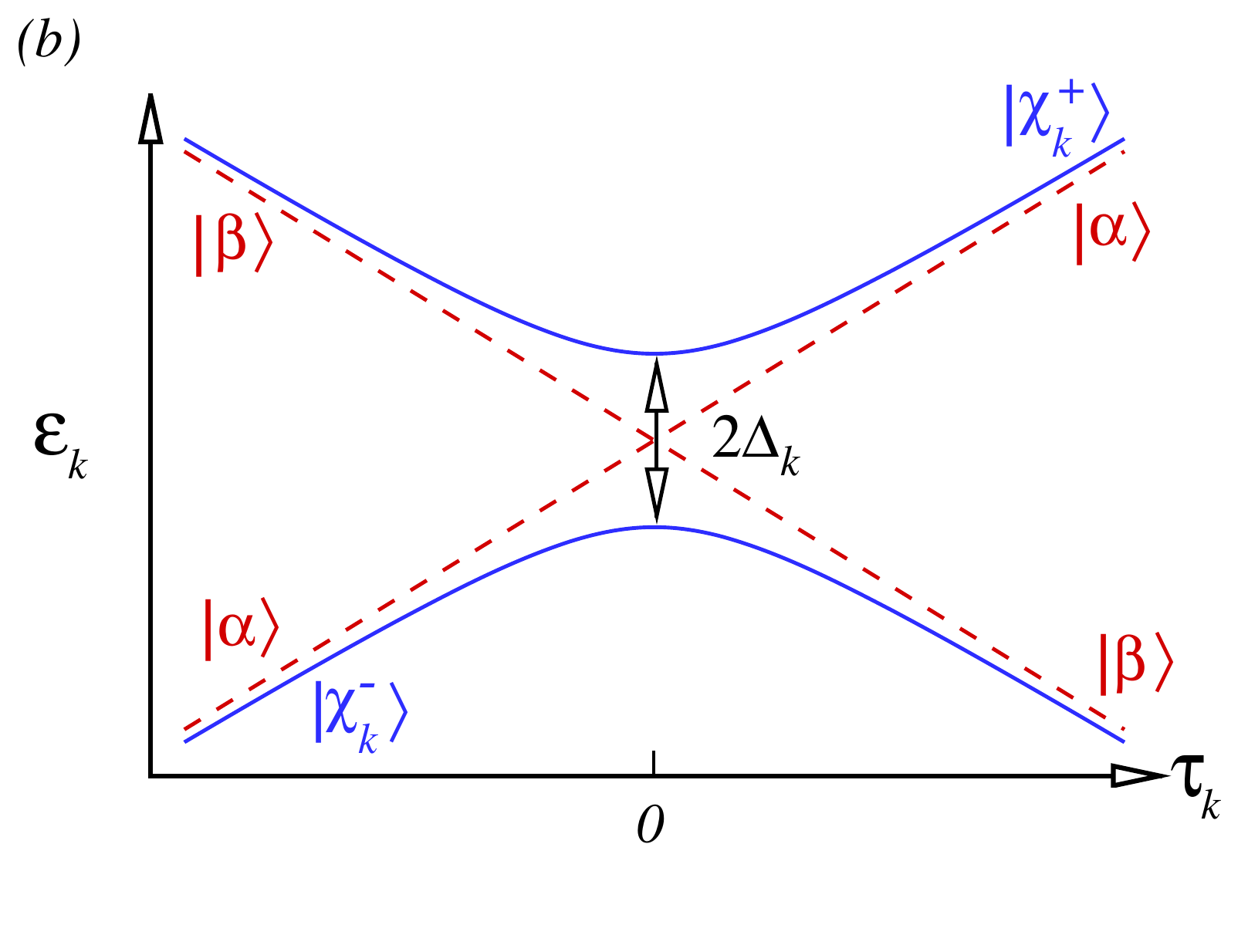}}
\centering
\end{minipage}
\caption{(Color online) (a) Illustration of a linear ramped quench (red color), with noise fluctuations superposed (grey color).
Here $h(t)$ is the magnetic field,
$h_i$ and $h_{\eta}$ its initial and final values, and $t_i$ and $t_{f}=0$ the corresponding times.
(b) Schematics of the instantaneous energies in the diabatic and adiabatic basis as a function of the effective mode-dependent time $\tau_k$
for the Hamiltonian, Eq. (3), in the main text \cite{Jafari2022bsm}. The diabatic energies are depicted by red dashed lines, with the adiabatic ones depicted by
blue lines, and with $2\Delta_{k}$ the gap between energy levels at $\tau_k=0$.}
\label{fig:LZ}
\end{figure*}
%

A straightforward way to calculate the critical times $t^{\ast}_n$ for a ramped quench takes off from the time evolution of the states $|\psi_k(\tau_{k,f})\rangle$ {\em after} the ramp. Bringing back the time variable $t$, and using that $t\!=\! t_{f}\!=\!0$ serves as reference time for the postquench dynamics, we introduce the notation $|\varphi_k(0)\rangle \equiv |\psi_k(\tau_{k,f})\rangle$, and write $u_k({\tau}_{k,f})\!\equiv\!u_k(t\!=\!0) \!=\! u_k(0)$ and
$v_k({\tau}_{k,f})\! \equiv \!v_k(t\!=\!0) \!=\! v_k(0)$. Further, we abbreviate the postquench Hamiltonian $H_{0,k}(0)$ as $H_{0,k}^{(f)}$. We are primarily interested in the Loschmidt amplitudes ${\cal G}_k(t)$ for the quasiparticle modes,
 %
\bea
\label{eq:Loschmidt}
{\cal G}_k(t)&=&\langle \varphi_k(0)|\exp(-iH_{0,k}^{(f)}\, t)| \varphi_k(0)\rangle\\
\no
&=& |u_k(0)|^2\exp(-i\varepsilon_{k,f}^+\, t) + |v_k(0)|^2\exp(-i \epsilon_{k,f}^-\, t),
\eea
%
with $\varepsilon_{k,f}^{\pm} = \varepsilon_k^{\pm}(0)$ defined in Eq. (5) in the main text \cite{Jafari2022bsm}. A DQPT is signaled by the vanishing of the Loschmidt amplitude
\begin{equation} \label{eq:fullLoschmidt}
{\cal G}(t) = \prod_k {\cal G}_k(t)
\end{equation}
 for the full system, causing a nonanalyticity in the rate function \cite{Heyl2013sm}
%
\bea
\label{eq:ratefunction1}
g(t) = - \lim_{N \rightarrow \infty} N^{-1}\ln |{\cal G}(t)|^2,
\eea
%
with $N$ the number of sites on the chain. The rate function $g(t)$ plays the role of a dynamical free energy density, with time $t$ standing in for a control parameter. Substituting $\prod_k {\cal G}_k(t)$ for ${\cal G}(t)$ in Eq. (\ref{eq:ratefunction1}), using $|v_k|^2=1-|u_k|^2$ and converting the product over $k$ into a sum, represented by an integral in the thermodynamic limit, one obtains
%
\bea
\label{eq:ratefunction2}
g(t) = \frac{-1}{2\pi} \int_{0}^{\pi} \log \Big(1 + 4 (|u_k(0)|^2-1) |u_k(0)|^2 \sin^2 (\frac{\varepsilon_{k,f}^+-\varepsilon_{k,f}^-}{2}) t \Big) dk.
\eea
%
Using that $\varepsilon_{k,f}^{\pm} = \pm \varepsilon_{k,f}$, the argument of the logarithm is seen to vanish with $g(t)$ becoming nonanalytic when $t=t^{\ast}_n$, where
%
\bea \label{eq:criticaltimes}
t_n^* = \frac{\pi} {\varepsilon_{k^{\ast}\!, f}}  \left(n+\frac{1}{2}\right), \ \ n=0,1,2,...
\label{eq:criticaltimes}
\eea
%
These are the critical times for the DQPTs, with $k^*$ the mode that satisfies $p_{k^{\ast}}\! =\! |u_{k^{\ast}}(0)|^2 \!=\! |v_{k^{\ast}}(0)|^2 \!=\! 1/2$, the existence of which was established above.

The examination of the condition for DQPTs following a noiseless ramp as reviewed here serves as a template when addressing the more intricate problem when noise is present during a ramp. This analysis is carried out in the main text \cite{Jafari2022bsm}.

\subsection{B. {\color{black} Transition probabilities in the presence of single noise realizations: \\ Stochastic Schr\"odinger equation}}

{\color{black} In the theory of stochastic differential equations {\color{black} (SDEs), the Ornstein-Uhlenbeck} (OU) process \cite{CoxMIller1965sm}, i.e., colored Gaussian noise $\eta(t)$ with zero mean $\langle \eta(t)\rangle=0$ and with (auto)correlation function
%
\begin{equation}
\label{OU}
{\color{black} \langle \eta(t)\eta(t')\rangle=\frac{\xi^{2}}{2\tau_n}e^{-|{t}-{t}'|/\tau_n},}
\end{equation}
%
can be generated from Gaussian white noise with zero mean $\langle \zeta({t})\rangle=0$ and correlation $\langle \zeta({t})\zeta({t}')\rangle=\xi^{2}\delta(t-t')$ through the SDE
%
\bea
\label{SDE}
\tau_n\dot{\eta}(t)=-\eta(t)+\zeta(t).
\eea
%
{\color{black} Here $\tau_n$ is the noise correlation time and $\xi$ the noise amplitude for fixed $\tau_n$.

In this work we have used the Mathematica built-in software {\em OrnsteinUhlenbeckProcess} to produce the OU noise. To obtain single
realizations of the continuous OU noise function $\eta(t)$ we interpolate the discrete points in the OU process with time step $dt = 0.01$.
With this, the transition probabilities $p_k$ (cf. Sec. A) in the presence of OU noise can be found} by numerically solving the stochastic Schr\"{o}dinger equations (SSEs) for a chain with $N$ sites,
%
\bea
\label{SSE}
i\frac{d}{dt}|\psi_k(t)\rangle=H_k(t)|\psi(t)\rangle=\Big(H_{0,k}(t)+\eta(t)H_{1}\Big)|\psi_k(t)\rangle, \ \ k = \frac{(2m-1)\pi}{N}, 
\eea
%
{\color{black} with $m=1,2,...,N/2$. Here $H_{0,k}(t)$ is defined in Eq. (3) of \cite{Jafari2022bsm} and $H_1 = 2\sigma^z$.

Our approach implies that the random noise variation is taken to be bounded at time intervals set by $dt$, reflecting that the physical system responds to noise with a finite time resolution.

For other numerical methods to solve SDEs with colored noise, see} \cite{Sancho1982sm,Fox1988sm,Billah1990sm,Mannella1992sm,dagpunar1988sm}.}

\subsection{C. {\color{black} Ensemble-averaged transition probabilities: Exact noise master equation}}

{\color{black} The solution of Eq. (\ref{SSE}) yields the transition probabilities $p_k$ given a single realization of the OU noise function $\eta(t)$. To obtain the {\em mean} transition probabilities
$-$ useful for uncovering features not easily seen from a single quench with a single noise realization$-$  one forms the ensemble averages $\langle p_k\rangle$ over the full noise distribution $\{\eta\}$.
In the following we outline the procedure how to go about this task.

For transparency and ease of notation,} we begin by considering a general time-dependent Hamiltonian,
%
\begin{equation} \label{Hamiltonian}
H(t)=H_{0}(t)+\eta(t)H_{1}(t),
\end{equation}
%
where $H_{0}(t)$ is noise-free while $H_{1}(t)$ is ``noisy" with $\eta(t)$ a real function for a given realization of the noise. This expression for $H(t)$, of the same structure as in Eq. (\ref{SSE}),
well captures linear corrections from a weak stochastic variation. As noted in Ref. \cite{Kiely2021sm}, the resulting formalism can readily be adapted to apply also beyond the linear regime.

{\color{black} As in Sec. B} we consider colored Gaussian noise $\eta(t)$ with mean $\langle \eta(t)\rangle=0$. The prototype form, {\color{black} OU} noise \cite{CoxMIller1965sm}
employed in the {\color{black} Letter} \cite{Jafari2022bsm}, is a {\color{black} stationary} stochastic process with {\color{black} (auto)correlation} function defined in (\ref{OU}).

{\color{black} With this setup} we now derive a noise master equation for the averaged density matrix $\rho(t)$ of $H(t)$ \cite{luczka1991quantumsm,Kiely2021sm}. One starts by writing down the von Neumann equation
%
\begin{equation}
\label{Neumann}
\dot{\rho}_{\eta}(t)=-i[H(t),\rho_{\eta}(t)],
\end{equation}
%
where
{\color{black}
\begin{equation} \label{quenchevolved}
\rho_{\eta}(t) = U^{\dagger}_{\eta}(t,t_i)\rho_{\eta}(t_i)U_{\eta}(t,t_i)
\end{equation}
 is the density matrix for a specific realization of the noise function $\eta(t)$. As follows from Eq. (\ref{Neumann}), the time-evolution from the noise-free initial condition $\rho_{\eta}(t_i)$ at time $t_i$ is carried through by $U_{\eta}(t,t_i) = {\cal T}\exp(-i\int_{t_i}^tH(t^{\prime})\, dt^{\prime})$, with ${\cal T}$ the time-ordering operator.}
 Introducing $\rho(t) = \langle \rho_\eta(t) \rangle$ as the ensemble average over many noise realizations (all with a common noise-free initial condition), the averaged von Neumann equation (\ref{Neumann}) takes the form
%
\begin{equation}
\label{meanNeumann}
\dot{\rho}(t)=-i[H_0(t),\rho(t)] - i[H_1(t),\langle \eta(t) \rho_\eta(t)\rangle].
\end{equation}
%
Applying a theorem by Novikov \cite{Novikov1965sm} one has that
%
\begin{equation}
\label{Novikov}
\langle \eta(t) \rho_\eta(t) \rangle = \langle \eta(t) \rangle \langle \rho_\eta(t) \rangle + \int_{t_i}^t \mbox{d}s \langle \eta(t) \eta(s) \rangle \langle \frac{\delta \rho_{\eta}}{\delta \eta}\rangle,
\end{equation}
%
with functional derivative
%
\begin{equation}
\label{func}
\frac{\delta \rho_{\eta}}{\delta \eta} = \frac{\partial\dot{\rho}_{\eta}}{\partial \eta} - \frac{\mbox{d}}{\mbox{d}t} \frac{\partial\dot{\rho}_{\eta}}{\partial\dot{\eta}}.
\end{equation}
%
Combining Eq. (\ref{func}) with (\ref{Hamiltonian}) and (\ref{Neumann}) gives
%
\begin{equation}
\label{func2}
\frac{\delta \rho_{\eta}}{\delta \eta} = -i[H_1(t),\eta(t)].
\end{equation}
%

The master equation follows by inserting Eq. (\ref{Novikov}) into (\ref{meanNeumann}), using Eqs. (\ref{OU}) and (\ref{func2}),
%
\begin{eqnarray}
\label{master}
\dot{\rho}(t)=-i[H_{0}(t),\rho(t)]-\frac{\xi^{2}}{2\tau_n}\Big[H_{1}(t),\int_{t_i}^{t}e^{-|t-s|/\tau_{n}}[H_{1}(t),\rho(s)]ds\Big].
\end{eqnarray}
%
The first term on the right-hand side accounts for the unitary time evolution generated by the prescheduled noiseless Hamiltonian
$H_{0}(t)$ and the second term induces the dynamics from {\color{black} OU} noise with Hamiltonian $H_{1}(t)$.\\

The 1D {\color{black} Jordan-Wigner transformed} quantum Ising Hamiltonian $H(t)$ with a noisy magnetic field studied in \cite{Jafari2022bsm} is expressed as a sum over decoupled mode Hamiltonians $H_k(t) = H_{0,k}(t) + \eta(t)H_1$, with $H_{0,k}(t)$ given in Eq. (3) and with $H_1=2\sigma^z$.
It follows that the density matrix $\rho_{\eta}(t)$ has a direct product structure \cite{Barouch1970sm}, i.e., $\rho_{\eta}(t)=\otimes_{k}\rho_{k,\eta}(t)$ with the {\color{black} $2\times2$ density matrix $\rho_{k,\eta}(t)$} satisfying $\dot{\rho}_{k,\eta}(t)=-i[H_{0,k}(t),\rho_{k,\eta}(t)]$ for a single common realization of the noise function $\eta$. {\color{black} $\rho_{k,\eta}(t)$ is here conveniently written in a rotating basis spanned by the instantaneous eigenstates $|\chi^{\pm}_k(t)\rangle$ of $H_k(t)$ (cf. Eq. (4) in \cite{Jafari2022bsm} with $\eta(t)$ added to $h_0(t)$). The noise master equation for the ensemble-averaged density matrix $\rho_{k}(t)=\langle\rho_{k,\eta}(t)\rangle$ takes the form}
%
\begin{equation}
\label{densitymode}
\dot{\rho}_k(t)=-i[H_{0,k}(t),\rho_k(t)]-\frac{\xi^{2}}{2\tau_n}\Big[H_{1},\int_{t_i}^{t}e^{-(t-s)/\tau_{n}}[H_{1},\rho_k(s)]\mbox{d}s\Big].
\end{equation}
%
%
\begin{figure*}
\begin{minipage}{\linewidth}
\centerline{\includegraphics[width=0.33\linewidth,height=0.25\linewidth]{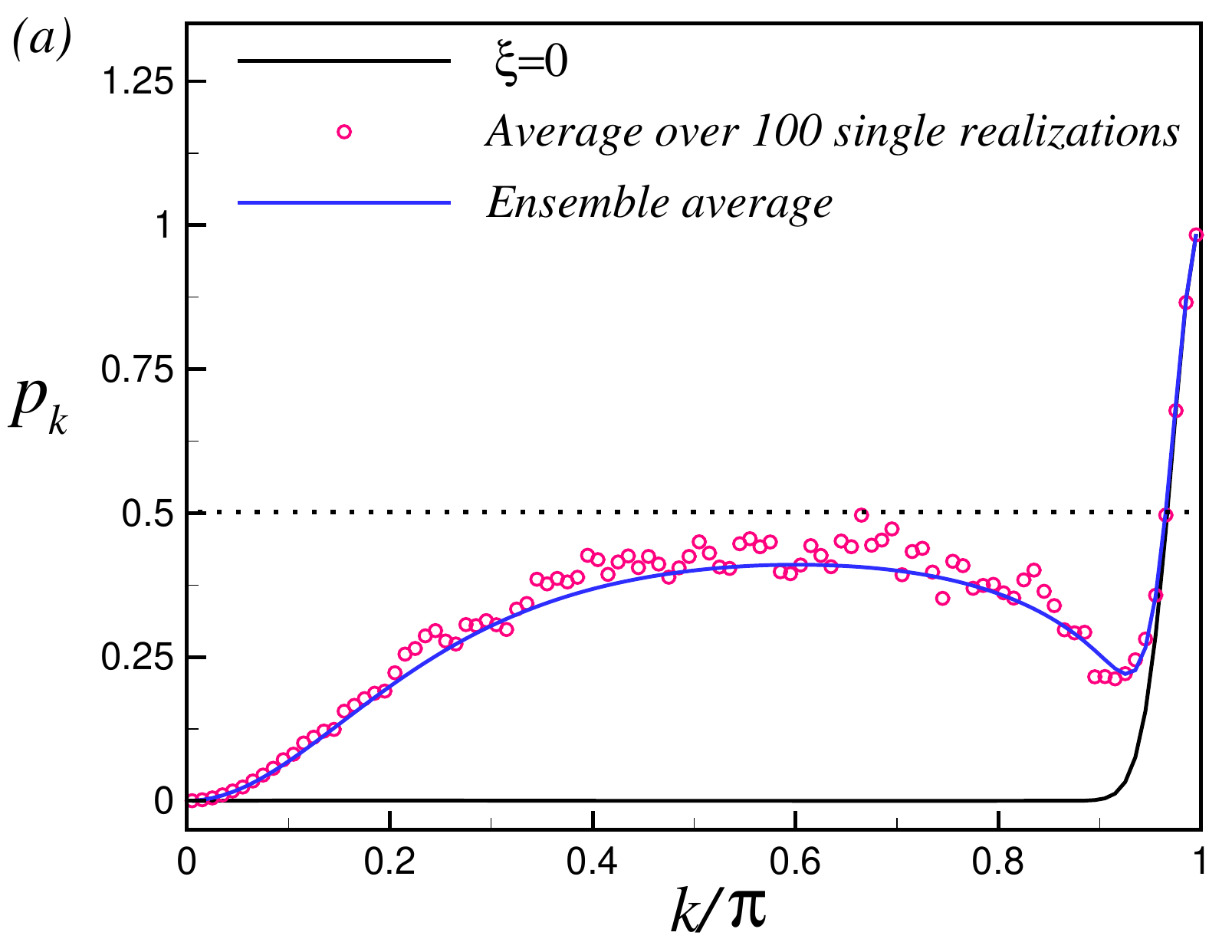}
\includegraphics[width=0.33\linewidth,height=0.25\linewidth]{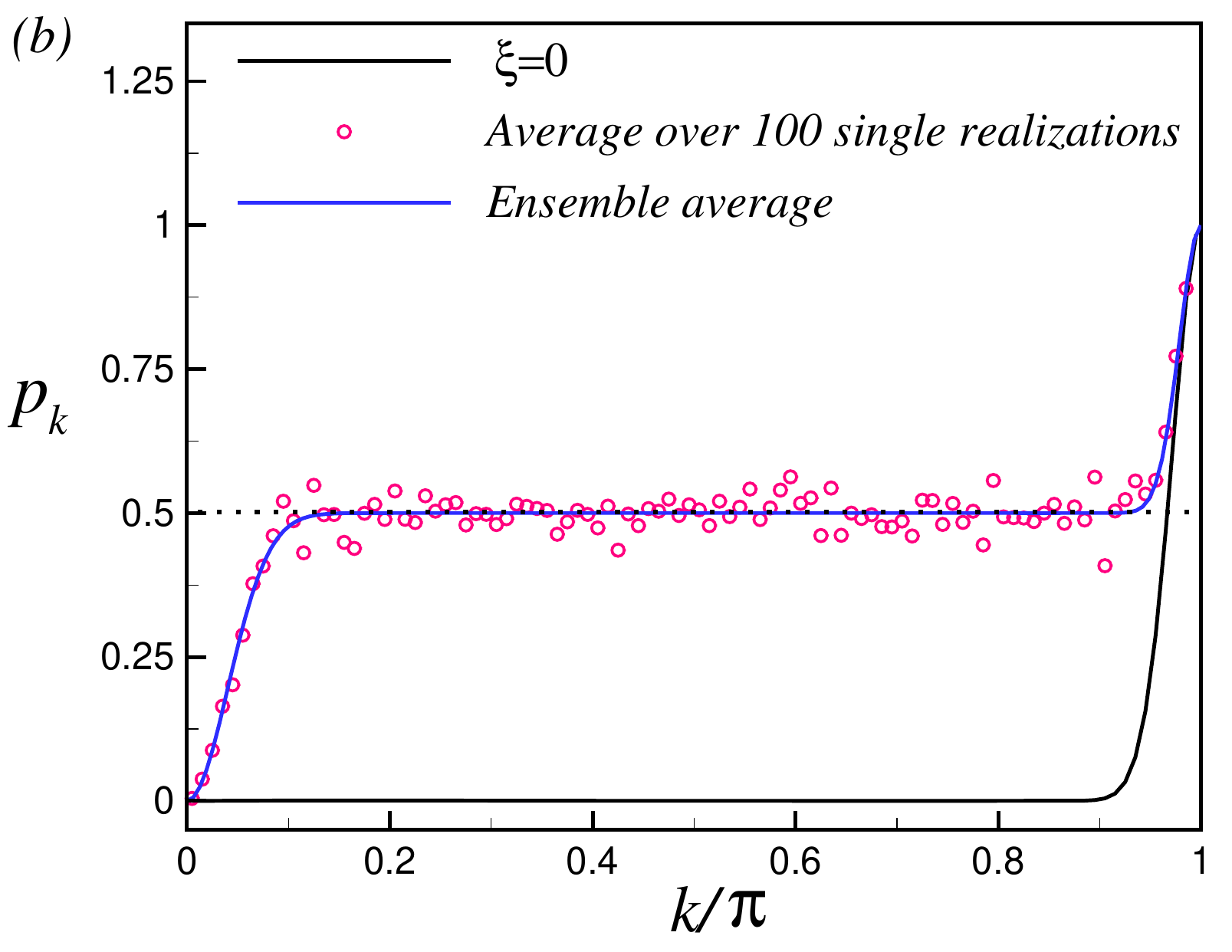}
\includegraphics[width=0.33\linewidth,height=0.25\linewidth]{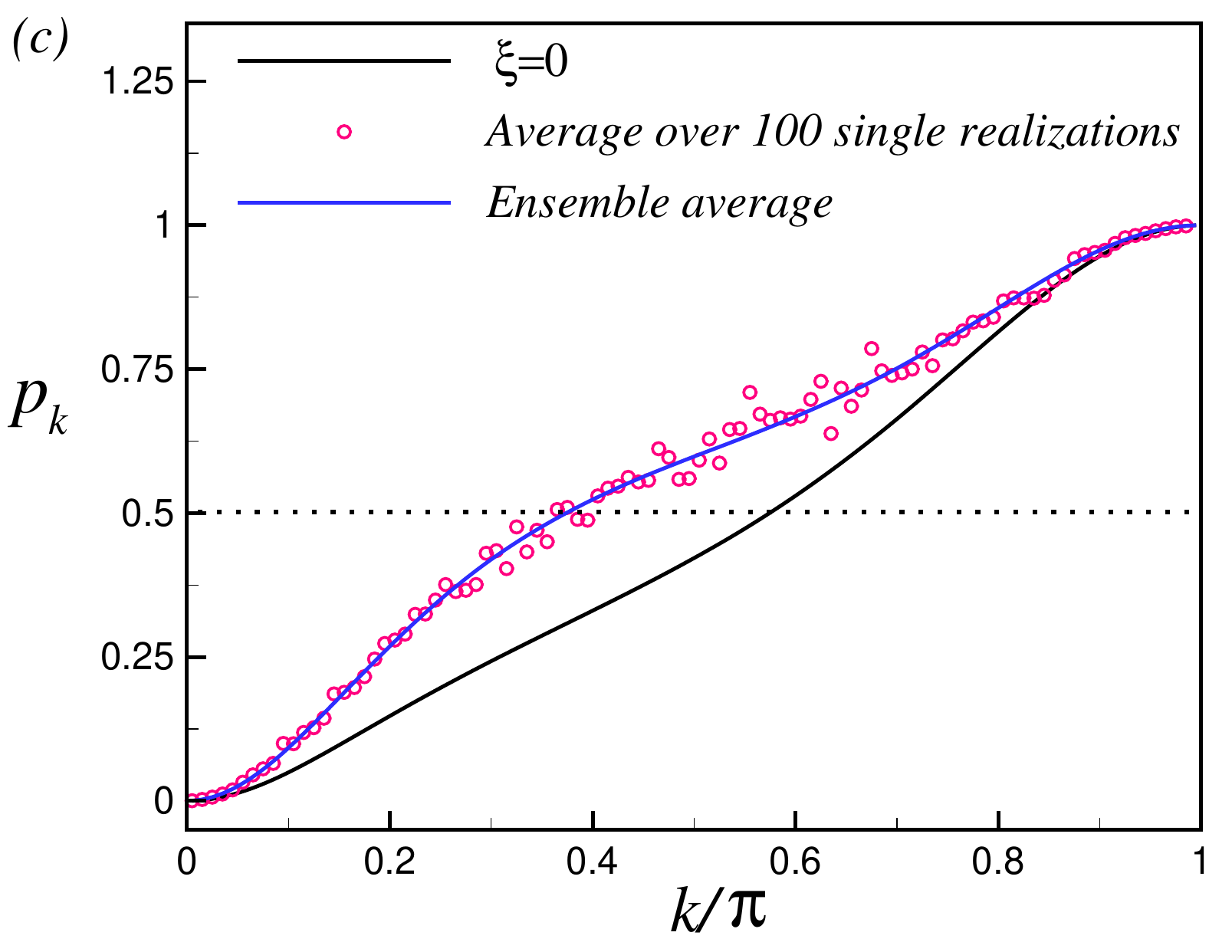}}
\centering
\end{minipage}
\caption{(Color online) {\color{black} Comparison of the average $p_k$ over 100 single realizations (red) and
the ensemble average (blue)  {\color{black} $\langle p_k \rangle$} for system size $N=200$ corresponding to {\color{black} the values of $\xi, v$, and $\tau_n$ as in} (a) Fig. 1(c); (b) Fig. 1(d) and (c) Fig. 1(f) in the main text \cite{Jafari2022bsm}. {\color{black} The probabilities for the noiseless cases $(\xi=0)$ are displayed in black.}}}
\label{figSMab}
\end{figure*}
%
To solve for $\rho_k(t)$, we consider the integral in the same equation as a new operator,
\bea
\label{DeltaOp}
\Gamma_k(t) \equiv \int_{t_i}^{t}e^{-(t-s)/\tau_{n}}[H_{1},\rho_k(s)]ds.
\eea
Eq. (\ref{densitymode}) then takes the form
%
\bea
\label{eqS15}
\dot{\rho}_k(t)=-i[H^{(0)}_k(t),\rho_k(t)]-\frac{\xi^{2}}{2\tau_n}[H_{1},\Gamma_k(t)].
\eea
%

By using the Leibniz integral rule, one obtains for the derivative of $\Gamma_k(t)$ with respect to time:
%
\bea
\label{eqS16}
\dot{\Gamma}_k(t)=-\Gamma_k(t)/\tau_n+[H_{1},\rho_k(t)].
\eea
%

The elements of the ensemble-averaged density matrix $\rho_k(t)$ can now be obtained by numerically solving the coupled differential equations (\ref{eqS15}) and (\ref{eqS16}) with initial
conditions
\begin{eqnarray} \label{initial1}
\rho_{k}(t_i)=
\left(\begin{array}{cc}
1 & 0 \\
0 & 0 \\
\end{array}
\right) \ \mbox{and} \ \
\Gamma_k(t_i)=
\left(\begin{array}{cc}
0 & 0 \\
0 & 0 \\
\end{array}
\right).
\end{eqnarray}
{\color{black} In the chosen basis, the first initial condition states that the system is initialized in the ground state, i.e., that all $k$-modes occupy the lower level $|\chi^{-}_k(t)\rangle$ at $t\!=\!t_i$.} The second condition simply expresses that the initial state is noiseless. {\color{black} With this, the ensemble-averaged nonadiabatic transition probability $\langle p_k \rangle$ for a mode $k$ is obtained as $\langle p_k \rangle = \rho_{k,22}(0)$, i.e., the ensemble-averaged probability that the $k$:th mode occupies the upper level $|\chi^{+}_k(0)\rangle$ at the end of the quench, $t\!=\!t_{f}\!=\!0$. Here recall that the choice of time reference serves as a reminder that $t_{f}=0$ is the {\em initial time} for the postquench dynamics.}

{\color{black} Having obtained the mean transition probability $\langle p_k \rangle$ for a mode $k$ from the master equation (\ref{densitymode}), we can use it to assess the validity of the solution to the SSE, Eq. (\ref{SSE}), for the same mode $k$: As illustrated} in Fig. \ref{figSMab}, the ensemble average $\langle p_k \rangle$ is clearly well reproduced by averaging over a finite sample of solutions $-$ {\color{black} each solution corresponding to a distinct single OU noise realization} $-$ to the corresponding SSE.

For detailed expositions of the approach to exact noise master equations, including formal properties of the time-evolved averaged density matrix, we refer the reader to Refs. \cite{luczka1991quantumsm,Kiely2021sm,Budini2000sm,Filho2017sm,Chenu2017sm}.

\end{widetext}

\end{document}